# On the theory of electric dc-conductivity: linear and non-linear microscopic evolution and macroscopic behaviour

J. Riess

*Centre de Recherches sur les Très Basses Températures, associé à l'Université Joseph Fourier, Centre National de la Recherche Scientifique, BP 166, 38042 Grenoble cedex 9, France*
*e-mail: jurg.riess@grenoble.cnrs.fr*

**Abstract**

We consider the Schrödinger time evolution of charged particles subject to a static substrate potential and to an homogeneous, macroscopic electric field (a magnetic field may also be present). We investigate the microscopic velocities and the resulting macroscopic current. We show that the microscopic velocities are in general non-linear with respect to the electric field. One kind of non-linearity arises from the highly non-linear adiabatic evolution and (or) from an admixture of parts of it in so-called intermediate states, and the other kind from non-quadratic transition rates between adiabatic states. The resulting macroscopic dc-current may or may not be linear in the field. Three cases can be distinguished : (a) The microscopic non-linearities can be neglected. This is assumed to be the case in linear response theory (Kubo formalism, ...). We give arguments which make it plausible that often such an assumption is indeed justified, in particular for the current parallel to the field. (b) The microscopic non-linearities lead to macroscopic non-linearities. An example is the onset of dissipation by increasing the electric field in the breakdown of the quantum Hall effect. (c) The macroscopic current is linear although the microscopic non-linearities constitute an essential part of it and cannot be neglected. We show that the Hall current of a quantized Hall plateau belongs to this case. This illustrates that macroscopic linearity does not necessarily result from microscopic linearity. In the second and third cases linear response theory is inadequate. We elucidate also some other problems related to linear response theory.

# 1. INTRODUCTION

This paper deals with the quantum mechanical time evolution of charged particles in a homogeneous electric field and the resulting macroscopic current. Most theories of electric conductivity are in some way or other based on a linearization with respect to the electric field. In particular linear response theory [1,2] is a systematic approach to calculate the linear response coefficients of the *macroscopic* current by linearizing the *microscopic*, quantum mechanical motion with respect to the electric field. The resulting general formulas are widely used for calculations of conductivities.

Nevertheless, this approach has been critizied by some authors. Van Kampen [3], from general arguments (partly based on classical examples) concluded that the microscopic motions are non-linear and therefore contested the well-foundness of linearizing the Schrödinger time evolution for the calculation of the linear macroscopic current. Lenstra and van Haeringen [4] analyzed a finite one-dimensional system in detail (a one-dimensional metal in a weak disorder potential) and observed that, *at sufficiently low electric fields,* the *microscopic, quantum mechanical* motion in this system is highly non-linear and that application of linear response formulas is only correct during unrealistically short time intervals, in which case the results have no meaning in relation to the conductivity problem. Further, based on explicit model calculations Riess [5] showed that microscopic non-linear velocity contributions are essential for the formation of the quantized Hall plateaus in the quantum Hall effect. Recently Wilkinson [6] examined the validity of the Kubo formula for electrical conductance of metallic systems by investigating a quantum model for dissipation by a random matrix method. He was able to identify a regime which exhibits Ohmic dissipation, but with a conductance which does not agree with the Kubo-Greenwood formula.

These findings question the general validity of linear response theory for the calculation of electric dc-conductivities. On the other hand, it is often considered to be an exact microscopic theory and it gives a satisfactory description of the conductivities in a vast number of experimental situations. However, it is *a priori* not clear in what cases linearizing the Schrödinger evolution is justified and when it is not.

In this paper we investigate this question qualitatively, in a way which is inspired by the discussion in [4], but from a more general point of view, and including situations with a magnetic field. Further we discuss some other questions raised by linear response theory.

We consider the time dependent Schrödinger equation of a system of charged particles in the presence of a homogeneous electric field **E** (called the macroscopic electric field). We call a *conducting state* a Schrödinger wave function whose velocity expectation value (averaged over a suitable small time interval) does not vanish.

In order to make our point as clear as possible we (first) consider a *system with discrete energy levels*, as e.g. in [2, 4]. For each individual state we investigate its time evolution which is induced by the macroscopic electric field **E**. For simplicity we consider a system of non-interacting particles with charge e (but our main results can be generalized to interacting particles, see later) and we neglect spin. We write the Hamiltonian in the representation

$$H(t) = (1/2)\left\{\left(\frac{\hbar\partial}{i\partial x}\right)^2 + \left[\frac{\hbar\partial}{i\partial y} - (e/c)\left[Bx + \phi(t)/L_y\right]\right]^2\right\} + (1/2m)\left(\frac{\hbar\partial}{i\partial z}\right)^2 + V(x,y,z) \quad (1)$$

where $\phi(t) = -cE_y L_y t$. and $V(x,y,z)$ is a static substrate potential. Without loss of generality we can choose the macroscopic electric field **E** = $(0,E_y,0)$ in y direction. In view of what follows below the magnetic field **B** = $(0,0,B)$ = curl$(0,Bx,0)$ is chosen in z-direction. Further, we chose a system of finite length $L_y$ in y-direction, with periodic boundary conditions in y-direction. The boundary conditions in x- and z- direction can be chosen arbitrarily as long as H is self-adjoint.

Proceeding as in [2, 4], we expand the solutions $\psi(x,y,z,t)$ of the time-dependent Schrödinger equation in terms of the momentary eigen functions $u_n(t)$ of the Hamiltonian H(t) (where t is considered as a parameter) :

$$\psi(t) = \sum_n c_n(t)u_n(t) \quad (2)$$

$$H(t)\psi_n(t) = \varepsilon_n(t)u_n(t). \quad (3)$$

Let us mention here that in the presence of a substrate potential without spatial symmetry (disorder) the one-parameter energies $\varepsilon_n(t)$ (generically) are indeed *non-intersecting* as a function of t (Wigner von Neumann anti-crossing theorem [7]). Further, due to the periodic boundary condition in the direction of the electric field, they are periodic in t with period $\tau = h/|eE_yL_y|$. see e. g. [8], as illustrated in Fig. 1. (Note, that $\tau$ is very small for realistic field values $E_y$, e. g., $\tau = 4\times10^{-12}$ s for $E_yL_y = 1$ mV). Therefore, for macroscopic samples the states $u_n(t)$ are insulating in the direction of **E**. This follows together with the general relation

$$\partial_t \langle\psi|H|\psi\rangle = e\mathbf{E}\mathbf{v}(t), \tag{4}$$

where $\mathbf{v}(t) = (v_x(t), v_y(t), v_z(t))$ is the expectation value of the velocity operator.

If $E_y$ is sufficiently small, the solutions $\psi(x,y,z,t)$ of the time dependent Schrödinger equation are *adiabatic* solutions [9], which coincide with the eigenfunctions $u_n(t)$ (apart from a possible factor of modulus unity). If $E_y$ increases, eventually any adiabatic state is modified by *non-adiabatic transitions* to other levels (Fig. 1), which means in general that the state becomes conducting in the direction of **E**. For sufficiently high electric fields (above a value which is characteristic for each state), a state is in the so-called *sudden approximation* [9]. Here the probability to follow an adiabatic time evolution has dropped to zero. In general, however, a state is *intermediate* between the adiabatic and the sudden approximation, i. e., a fraction of the wave function follows adiabatic evolutions and a fraction makes non-adiabatic transitions (Fig. 1).

## 2. PARALLEL CURRENT

Let us calculate the current of the system resulting from the Schrödinger time evolution in the presence of **E**. First we consider the *current parallel to **E***. In our case this is the current in y direction. Due to (2) - (4), summing over the velocities of all the particles (and neglecting shot noise), the total electric current in y direction can be written as (omitting spin degeneracy)

$$I_y(t) = (e/L_y) \sum_{j}^{N} v_y^j(t) = [1/(E_y L_y)] \left\{ \sum_{j,n} |c_n^j(t)|^2 \partial_t \varepsilon_n(t) + \sum_{j,n} \varepsilon_n(t) \partial_t |c_n^j(t)|^2 \right\} \quad (5)$$

here N is the number of particles in the system (j = 1, 2, ..., N). The first term on the right hand side of (5) is the sum of the currents of the adiabatic states with the momentary occupation numbers

$$\rho_{nn}(t) = \sum_{j}^{N} c_n^{*j}(t) c_n^j(t). \quad (6)$$

The second term is due to transitions between the levels $\varepsilon_n(t)$. Eq. (5) corresponds to Eq. (16) of [10] or to (11) of [4] where the second term is written in a different form, which also can be written as

$$\sum_{n,n';n \neq n'} (2e/L_y) \rho_{nn'}(t) (v_y^{op})_{n'n} \quad (5')$$

[here spin degeneracy has been included and $v_y^{op} = (-i\hbar/m)\partial/\partial y$].

For *macroscopic* purposes we can average expression (5) over the time period $\tau$ defined above, since $\tau$ is generally very small and tends to zero for infinite system size. Because of the periodicity of the levels $\varepsilon_n(t)$, the first term on the right hand side of (5) does not contribute to this macroscopic (i. e., $\tau$-averaged) current $I_y(t)$ if the occupation numbers $\rho_{nn}(t)$ are constant in time. Let us now consider the general case, where $\rho_{nn}(t)$ does depend on t and evaluate the contribution of a state which fully occupies the level $\varepsilon_i(t)$ at the initial time t, i. e., where we have the initial conditions

$$|c_i(t)|^2 = 1, |c_n(t)|^2 = 0 \text{ for } n \neq i.$$

As a consequence of (4) the average velocity of this state over the time interval $\tau$ is

$$\{<H(t+\tau)> - <H(t)>\}/(\tau e E_y) = \sum_n [\,|c_n(t+\tau)|^2 - |c_n(t)|^2\,]\, \varepsilon_n/(\tau e E_y)$$

$$= [\sum_{n \neq i} \varepsilon_n |c_n(t+\tau)|^2 + \varepsilon_i \{|c_i(t+\tau)|^2 - |c_i(t)|^2\}]/(\tau e E_y). \qquad (7)$$

Since the term in the last curly bracket is equal to $-\sum_{n \neq i} |c_n(t+\tau)|^2$ expression (7) finally becomes $\sum_{n \neq i} |c_n(t+\tau)|^2 [\varepsilon_n - \varepsilon_i]/(\tau e E_y)$, where $|c_n(t+\tau)|^2$ is just the probability to make a transitions from $\varepsilon_j$ to $\varepsilon_n$ in the considered time interval $\tau$. The remaining probability $|c_i(t+\tau)|^2$ to stay on the level $\varepsilon_i$ contributes nothing to (7) since $\varepsilon_i(t+\tau) = \varepsilon_i(t)$. This means that the macroscopic parallel current is indeed generated by the transitions between the levels $\varepsilon_n$.

From this one might conclude that the adiabatic portions of the total time evolution of a state do not contribute to the macroscopic parallel current. However, beyond a time interval of length $\tau/2$, the evolution from the initial level $\varepsilon_i(t)$ to other levels may contain pieces of adiabatic paths and thus make the corresponding evolutions to other levels non-linear. For the parallel conductivity these non-linearities can be neglected in first aproximation if, during the time between successive collisions with the surroundings (phonons), the non-adiabatic transition probabilities for the states near the Fermi level are sensibly larger than the probabilities for the adiabatic evolution (which is the typical case in metals). This situation is favoured in cases where non-adiabatic transitions simultaneously occur to many other levels (which in turn is favoured if the spectrum tends to a continuum).

### 3. LINEARIZATION

The linear response expressions for the current (5) (eq. (29) of ref. [2] or eq. (19) of ref. [4]) are based on the linearized equation for the density matrix in a treatment which essentially neglects the adiabatic evolution (see also sec. 5 of ref. [4]). This means they correspond to the *second term* of equation (5) in linear approximation with respect to the electric field **E**, i. e., to the *linearized* non-adiabatic transitions or, in view of eq.

(4), to quadratic non-adiabatic transition rates. (Further, in order to get a time independent current, the quadratic transition rates must become time independent).

Due to this equivalence the linearized macroscopic parallel current is sometimes calculated directly by the transition rates in second order between *stationary* levels $\varepsilon_n(0)$, using *Fermi's golden rule* formula, see e. g. [11, 12], which indeed leads to a t-independent result. (Incidentally, this implies that the occupation numbers of the states $u_n(t=0)$ represent the *steady state in such a theory*, see below.)

One might now expect that in cases where only the non-adiabatic transitions are important, the linearized parallel current obtained with Fermi's golden rule gives physically sound results. However, there is a well known problem with this formula, since *for a constant homogeneous electric field* it gives zero in the case of discrete levels. Non-zero results are only obtained for a continuum (for discrete levels non-zero results can however be obtained by introducing ad hoc a broadening of the levels which is justified as being the result of the interaction with a heat bath [12]). This is surprising since explicit calculations for systems with discrete levels which are based on the entire time evolution with all non-linearities included, generally lead to a non-vanishing parallel current even for a constant, homogeneous field **E**, and the current can be time-independent and linear in the field, see e. g. [4, 10].

The reason for this discrepancy can be elucidated by a more general formula for the non-adiabatic transition probability $p_{nm}$ to go from the discrete level $\varepsilon_n(t)$ to another discrete level $\varepsilon_m(t)$ in the time interval from $t_0$ to $t_1$, given by eq. (XVII.111) of ref. [9]:

$$p_{nm} \approx |\int_{t_0}^{t_1} \alpha_{mn}(t) \exp[i\int_{t_0}^{t} \omega_{mn}(\tau)d\tau]dt|^2 \qquad (8)$$

Here $\omega_{mn}(t) = (\varepsilon_m(t) - \varepsilon_n(t))/\hbar$, and $\alpha_{mn}(t) = <u_m|du_n/dt>$, which is a measure for the speed of change of $u_n(t)$ as a function of time. (Expression (8) results from the first term of the iterative perturbation expansion for the operator W, defined by (XVII.87) of Ref. 9, where W determines the deviation of the time evolution operator from the adiabatic evolution.) In the case of *discrete*, *stationary* states relation (8) gives zero as Fermi's golden rule does. (We remark however that the derivation of Fermi's golden rule needs a transition time interval which is sufficiently large, whereas the more general fomula (8) is valid for a transition between arbitray times $t_0$ andt $t_1$).

Now in general the states $u_n(t)$ are *not stationary*, i. e., they do *depend on t, even in the presence of a constant, time-independent electric* field (see below). It is because of this time-dependence (which is apparent in formula (8)) that the non-adiabatic transition probabilities are in general different from zero also in the case of discrete levels. Therefore, on general physical grounds it is not necessary to have a continuum in order to get non-zero values of the parallel dc current. A continuum is only required for the linear approximation.

In the weak disorder case the Landau-Zener formula gives an explicit expression for the transition probability $p_{n,n\pm1}$ to a next nearest level in the time interval $\tau/2$ (e. g., from $\varepsilon_n(-0,25)$ to $\varepsilon_{n+1}(0,25)$ in Fig. 2a). Here the electric field dependence is of the form $p_{n,n\pm1} = \exp(-|\text{constant}/E_y|)$, which goes to zero faster than any power of $E_y$ when $E_y$ goes to zero (adiabatic approximation) and tends to 1 as $1 - |\text{constant}/E_y|$ when $E_y$ goes to « infinity » (sudden approximation). Unfortunately, no explicit formula exists in the general case. Nevertheless, one can show (see Appendix A) that the transition rates $dp_{nm}/dt$ are of the form of the Fermi golden rule (hence are indeed quadratic in $E_y$) under the following conditions. The first condition is that the $p_{nm}$ are described by the approximate relation (8). Further, $\alpha_{mn}$, $\varepsilon_n$ and $\varepsilon_m$ must be constant in time. In the following we will give an argument which makes it plausible that in the continuum limit (i. e., when the system is sufficiently large) the fulfilment of the last two conditions is favored. This seems then to explain why in many cases the linear response formulas are good approximations when applied to a continuum, at least for experimentally « reasonable » electric field values.

To this end we first consider the case of a *one-dimensional* system of finite length L (= $L_y$), with V(y) an asymmetric potential (disorder) and B = 0. Further, we start by choosing the physical parameters such that the system can be described by a weak disorder approximation, because in this case the transition rates are generally different from beeing quadratic. Later the continuum will be approached by L tending towards infinity. Fig. 2a illustrates the behaviour of the discrete energy levels in the weak disorder case (see e. g. [4, 10]). Here an adiabatic function $u_n(t)$ is a linear combination of at *most two* basis functions at a given time. In our one dimensional example the basis functions are the solutions in the absence of V(y), i. e., the free particle functions $\exp(iky)/L^{1/2}$, where $k = 2\pi p/L$, p integer, and the energy gap $g_{k,k'}$ between $\varepsilon_n$ and $\varepsilon_{n'}$ in Fig. 2a has the form

$$g_{k,k'} = |(2/L)\int_0^L V(y)\exp[i(k-k')y]dy|.$$

The weak disorder approximation is valid if $g_{k,k'} << |E_k(\tau/2) - E_{k'}(\tau/2)|$.

Within the small transition intervals $\Delta_{kk'}$, which approximately is given by the time interval where the adiabatic energies $\varepsilon_n$ and $\varepsilon_{n+1}$ deviate from the free-particle energies $E_k(t)$ or $E_{k'}(t)$, the adiabatic function $u_n(t)$ changes from practically a plane wave $\exp(iky)/L^{1/2}$ to $\exp(ik'y)/L^{1/2}$, while $u_{n+1}(t)$ changes from $\exp(ik'y)/L^{1/2}$ to $\exp(iky)/L^{1/2}$. This means that the absolute value of $\alpha_{n+1n}(t)$ qualitatively behaves as in Fig. 3a.

When L increases, more and more unperturbed levels (the dashed lines in Fig. 2) appear between the initially considered levels, giving rise to new adiabatic levels with their anti-crossings and their corresponding transition time intervals. As an example let us consider the simple case when the original system of length L is repeated once to obtain a new system of length 2L (Figs. 2b and 3b). Here $g_{kk'}$ and $\Delta_{kk'}$ remain unchanged if in Fig. 2b we consider the adiabatic functions $\varepsilon'_m$ and $\varepsilon'_{m+1}$ made of the two plane waves with the same k vectors as $\varepsilon_n$ and $\varepsilon_{n+1}$ in Fig. 2a. Note that in Figs. 2b, 3b the value of $\tau$ is half of that in Figs. 2a, 3a. Therefore the transition interval $\Delta_{kk'}$ comes now closer to its neighbouring transition intervals.

When L further increases (hence $\tau$ diminishes), again new levels appear and eventually neigbouring transition intervals will touch and finally overlap each other. Qualitatively this means that for sufficiently large L the interval $\tau/2$ is contained in an interval where $\alpha_{n+1n}(t)$ *is practically constant*. In this limit of overlapping « weak disorder transition intervals » the weak disorder approximation has broken down and we are in a strong disorder scenario.

In this situation, which one expects to occur quite generally when the system length in the direction of **E** is sufficiently large, an adiabatic function $u_n(t)$ is a linear combination of more than two plane waves at a given time t and its change in the time intervals of length $\tau/2$ is no longer a change between two entire plane waves $\psi_k$ and $\psi_{k'}$ (corresponding to the behaviour of $\alpha_{n+1\,n}$ in Fig. 3) but is now a much slower and less fluctuating change, leading to almost constant $\alpha_{n'n}(t)$ with time.

If we add one or two additional dimensions to the original one, then new anti-crossings occur already in the original picture of Fig. 2a with the original system length L. This means that the strong disorder scenario is reached already for a smaller length L.

In general the change of a function $u_n(t)$ can be visualized by the motion in space of its phase structure represented by the lines of constant phase function. These lines converge to phase singularities, which in three dimensions are nodal lines of $u_n(t)$ with non-zero phase winding numbers (see Appendix B and [8, 13]). For a function in the weak disorder case the phase singularities move very rapidly during a *fraction* of the interval $\tau/2$ but practically do not move during the rest of this time interval [8, 13], resulting in the qualitative behaviour of $\alpha_{n'n}(t)$ shown in Fig. 3. On the other hand, functions in the strong disorder case show a *slow* but continuous motion of phase singularities during the *entire* interval $\tau/2$, resulting in *almost constant* $\alpha_{n'n}(t)$.

In summary, we have made plausible that in the presence of a disorder substrate potential a sufficiently large system always tends to a strong disorder situation and that here not only the energy levels $\varepsilon_n(t)$ but also the corresponding expressions $\alpha_{n'n}(t)$ tend to be constant (while the functions $u_n(t)$ are not constant). As we have mentioned above, in this case the transition rates $dp_{nn'}/dt$ derived from equation (8) formally tend to Fermi's golden rule (for sufficiently large t), i. e., indeed to an expression quadratic in $E_y$ and independent of t. However, one has to keep in mind that equation (8) is based on the first order iteration term of the operator W mentioned above and may not be a good approximation for all values of $E_y$. This may in particular be the case for very small or very high values of $E_y$, as is suggested by the Landau-Zener probability in the weak disorder case, see above.

As we have seen, linear response theory for the parallel conductivity is based on the quadratic part of the non-adiabatic transition rates (or equivalently, on the quadratic part of the energy change per unit time in the representation (1)). In cases where these quadratic parts are missing or are not dominant, the mentioned linear response formalism is inadequate (e. g. for free particles).

The parallel conductivity is associated with *dissipation*. The non-adiabatic transitions caused by the electric field in the vicinity of the Fermi level change the occupation numbers $f_n$ (= $\rho_{nn}$) of the levels $\varepsilon_n$ and thus drive the system out of its thermodynamic equilibrium. Simultaneously, the interaction with the surrounding heat bath (phonons) at temperature T tends to bring the occupation numbers $f_n$ back towards their equilibrium value. These two counteracting mechanisms lead to a steady state distribution $f_n^{steady}$ which is different from the Fermi distribution.

However, in the linear response approximation of the current the steady state distribution does not appear and is replaced by the equlibrium distribution $f_n^{Fermi}$. To see why this turns out in this way we write the parallel current in the steady state in terms of non-adiabatic transition rates $W_{nm}^{Sch}$ = $dp_{nm}/dt$ (from the Schrödinger time evolution)

$$I_y^{steady} = \sum_{nm} (\varepsilon_m - \varepsilon_n) W_{nm}^{Sch} f_n^{steady} [1 - f_m^{steady}]/(E_y L_y). \tag{9}$$

In the linear approximation for the current $W_{nm}^{Sch}$ is quadratic in $E_y$ as we have seen, and with $W_{nm}^{Sch} = W_{mn}^{Sch}$ (which follows e. g. from the form of $p_{nm}$ given in Appendix A) the sum of products $f_n^{steady} f_m^{steady}$ in (9) vanishes, i. e., the last bracket can be omitted. If we write $f_n^{steady} = f_n^{Fermi} + a_n E_y$ + higher orders in $E_y$, we see that a linear expression for $I_y^{steady}$ must contain $f_n^{steady}$ in the approximation $f_n^{Fermi}$.

Since, as a consequence of linearization, the equilibrium distribution $f_n^{Fermi}$ (i. e., the distribution before switching on the electric field) is used instead of the steady state distribution, the mechanism of dissipation, i. e., the interaction with the phonons, does not appear in linear response theory and is here irrelevant. However, it has to be *postulated* that a steady state *exists* (although it is not calculated). In this sense linear response theory is phenomenological [14]. Further, the deviation of $f_n^{steady}$ from $f_n^{Fermi}$ must assumed to be small.

In the steady state distribution the total rate of energy change due to the Schrödinger time evolution caused by **E** is exactly compensated (annihilated) by the rate of energy change caused by the interaction with the phonon bath. For each particle this interaction is supposed to take place at its momentary position (in **E**-direction) at time t. In cases, where the steady state distribution is effectively calculated, the representation (1) of H is well adapted since here the energies are *automatically* measured with respect to the chemical potential at the momentary position of each particle. Further, in case of periodic boundary conditions in the direction of **E** the representation (1) has to be used also for mathematical reasons, because in the representation where the electric field appears as a static potential -e**Er**, the Hamiltonian would not be a self-adjoint operator.

The steady state depends on the transition rates generated by the interaction with the surrounding heat bath (phonons). They determine the average time between two scattering events of a state of given energy with the phonons. For conducting states near the Fermi level this time can be sufficiently short such that only one or a few adiabatic half oscillation periods can develop and the role of adiabatic velocity contributions tends to be small or negligeable. This is relevant for the parallel current since here only non-adiabatic transitions among states near the Fermi level contribute (contributions from lower lying states cancel). On the other hand, for states sufficiently below the Fermi level this scattering time tends to be infinite. This is important for the *Hall current* $I_x$. Here the contributions from states below the Fermi level do not cancel and therefore the adiabatic velocity contributions cannot in general be neglected, especially in high magnetic fields.

### 4. HALL CURRENT

To see this in more detail we consider the Hamiltonian (1), again with periodic bounday conditions in y-direction. Further we assume that $|\Psi| \to 0$ if $|x|$ goes to infinity such that the position operator x becomes a self-adjoint operator. (For simplicity, and also in view of an application to the quantum Hall effect, we neglect the extension in z-direction). The Hall velocity, i. e., the velocity in x-direction of a state $\Psi$ can then be written as $\partial_t <\psi|x|\psi>$.

From Appendix B we know that the modulus of a function $u_n(\mathbf{r},t,E_y)$ is periodic in t with period $\tau$ (for a given $E_y$) and periodic in $E_y$ with period $h/(eL_y t)$ for a given t and further, that $|u_n(\mathbf{r},t,E_y)|$ is neither constant in t nor constant in $E_y$. This leads to a genuine oscillation of the modulus with respect to t and $E_y$ which in general leads to a non-vanishing, oscillating contribution to the corresponding adiabatic velocity, and in particular to a non-vanishing component $v_x = \partial_t <u_n|x|u_n>$.

Let us now consider a state $\psi$ in the representation (2), again with the initial conditions

$$|c_i(t=0)|^2 = 1, |c_n(t=0)|^2 = 0 \text{ for } n \neq i.$$

The macroscopic (i. e., $\tau$ averaged) Hall velocity $v_x$ associated with $\psi$ is different from zero if $|c_i(t)|^2$ changes with time, i. e. if the evolution is not pure adiabatic. This happens in particular if $\psi$ is a state intermediate between the sudden and the adiabatic approximation. *The adiabatic contributions to the Hall velocity of an intermediate state cannot in general be neglected.*

To illustrate this, we consider a weak disorder model [15, 5, 16], since here the time evolution becomes particularly transparent. Fig. 4 schematically reproduces the corresponding time evolution in the energy time plane. At $t = \tau/2$ the initial function $\psi(t=0) = u_i(t=0)$ has developed into $c_i(\tau/2)u_i(\tau/2) + c_{i+1}(\tau/2)u_{i+1}(\tau/2)$ (for details of the calculation see the quoted references). The first term represents the result of the adiabatic evolution (from 1 to 4) and the second term the result of the *non-adiabatic* transition (from 1 to 2)**.**

For our qualitative discussion of the associated velocities it is sufficient to make the simplifying approximation that the transition times into different branches are zero, i. e., that the wave function undergoes « splittings » at the discrete times $t = \tau/4$ modulo

τ/2. The contribution to the Hall velocity due to the transition from 1 to 2, averaged over the interval $0 \leq t \leq \tau/2$, becomes then $|c_{i+1}(\tau/2)|^2[<u_{i+1}(\tau/2)|x|u_{i+1}(\tau/2)> - <u_i(0)|x|u_i(0)>]/(\tau/2) = |c_{i+1}(\tau/2)|^2 cE_y/B$ ([5,15,16]). (Incidentally, this transition from 1 to 2 represents the sudden approximation branch of ψ in the time interval $0 \leq t \leq \tau/2$ and the equality sign illustrates the general result [17, 18] that the Hall velocity of a function in the sudden approximation, i. e., which does not contain any adiabatic evolution, is equal to the unperturbed classical Hall velocity.)

In the same time interval of length τ/2 the fraction $|c_i(\tau/2)|^2$ of ψ follows the *adiabatic* evolution from 1 to 4, which gives rise to the average Hall velocity contribution

$$|c_i(\tau/2)|^2[<u_i(\tau/2)|x|u_i(\tau/2)> - <u_i(0)|x|u_i(0)>]/(\tau/2).$$

This is a non-classical, non-vanishing velocity part due to the change of the shape of $|u_i(\mathbf{r},t)|^2$ in this time interval. Therefore the adiabatic evolution from 1 to 4 does contribute to the average *Hall* velocity $v_x$, while it does not contribute to the *parallel* velocity $v_y$. The latter is only generated by the non-adiabatic branch from 1 to 2 and gives $v_y = 2|c_{i+1}(\tau/2)|^2[\varepsilon_{i+1}(\tau/2) - \varepsilon_i(0)]/(\tau eE_y)$, in view of (4). We emphasize that the contribution to the Hall velocity due to the adiabatic (non-classical) evolution from 1 to 4 can be much bigger than the corresponding non-adiabatic (classical) contribution from 1 to 2.

Two adiabatic paths over consecutive intervals of length τ/2, e. g., from 1 to 4 and from 4 to 6 in Fig. 4 or from 1 to 2 and 2 to 3 in Fig. 5, individually have zero average *parallel* velocity $v_y$. On the other hand, their corresponding *Hall velocities* are not zero individually, but in a pure adiabatic state $u_i(\mathbf{r},t)$ they would be opposite equal because of the τ periodicity of $|u_i(\mathbf{r},t)|^2$ and therefore cancel each other. Such a cancellation does no longer happen in an *intermediate state,* i. e., in a state which is neither in the adiabatic nor in the sudden approximation, since here the time evolution in general attributes *different probabilities to such consecutive adiabatic paths.* Therefore, the adiabatic, i. e., non-linear contributions to the total Hall current cannot in general be neglected if intermediate states are present.

In the quantum Hall effect the intermediate states of a broadened Landau band are sandwiched between adiabatic (i. e. insulating) states at the upper and lower mobility edges and the sudden approximation states in the center of the band. Further, the values of the probabilities associated with the adiabatic branches of the intermediate states follow a general scenario, which we illustrate schematically in Fig. 5 by our weak disorder example.

Here, in the energy range between the two mobility edges $\varepsilon^+$ and $\varepsilon^-$ the following holds. In the lower half the probabilities $P_{jk}$ to follow an *adiabatic* path from j to k in Fig. 5 decrease from the lower mobility edge $\varepsilon^-$ towards the band center in the following way :

$1 = P_{12} > P_{23} = P_{56} > P_{45} = P_{78} > P_{89}$ and so on, and in the upper half we have $1 = P_{1'2'} > P_{2'3'} = P_{5'6'} > P_{4'5'} = P_{7'8'} > P_{8'9'}$ and so on. This is related to particle conservation and to the general fact that the *non-adiabatic* transition probabilities decrease from the band center towards the two mobility edges (where they are zero). In our model we have $P_{26} = P_{53} < P_{48} = P_{75} < ...$ and so on, and $P_{2'6'} = P_{5'3'} < P_{4'8'} = P_{7'5'} < ....$ and so on.

Further, *in the lower half of the band* in Fig. 5 the Hall velocities of the adiabatic paths with downward (upward) curvature, as e. g. from 1 to 2 (2 to 3), have the same direction as those generated by adiabatic paths with upward (downward) curvature in the upper half, as e. g. from 1' to 2' (2' to 3'). In addition, the Hall velocities associated with downward (upward) curvature paths in the lower half, or with upward (downward) curvature paths in the upper half have the same (opposite) direction as the Hall velocities associated with the non-adiabatic paths. The latter are all in the direction of $cE_y/B$.

From the fact that the Hall velocities (averaged over $\tau/2$) of consecutive adiabatic paths of length $\tau/2$ (as e. g., from 1 to 2 and from 2 to 3) are opposite equal, together with the law for the probabilities of the adiabatic branches mentioned above, it follows then that the *sum* of all the *adiabatic* (hence *non-linear*) *Hall* velocity contributions of the intermediate states in a broadened Landau band *is different from zero*.

Since the *Hall* velocities of the *non-adiabatic* paths (going up or downwards in the energy time plane) all are in the direction of $cE_y/B$, they do not cancel each other, in contrast to the corresponding *parallel* velocities. Thus in a fully occupied broadened Landau band, or when the Fermi level is in a range of insulating (i. e., adiabatic) states above the upper mobility edge, the sum of the parallel velocities vanishes, whereas the sum of the Hall velocities leads to a non-zero macroscopic Hall current (provided not all occupied states are adiabatic). This current contains contributions from non-adiabatic and adiabatic paths. The latter are non-linear in $E_y$, as we have seen.

The relevance for the quantum Hall effect of microscopic non-linear velocity contributions can also be shown by a formal proof which is independent of any particular model. Typically, the *integer* quantum Hall effect is due to the presence of insulating states in the tails and conducting states in the centre of the disorder broadened Landau bands. When the Fermi level is contained in the range of insulating states above a band center and the Hall conductivity is quantized, the few conducting states in the centre of the band together carry the same Hall current as a fully occupied, unperturbed Landau band, i. e., the sum of all the Hall velocities is equal to $NcE_y/B$, where $N = L_xL_yeB/hc$ is the number of states per Landau band in the area $L_xL_y$. This means that the conducting states compensate the missing current of the insulating states in the tails (one speaks of *compensating current*). Since N is bigger than the number of conducting states in the band, this implies that there exist conducting states whose Hall velocities are higher than the unperturbed, classical Hall velocity $cE_y/B$.

Now the following general result is important. A state which is in the sudden approximation has the same Hall velocity $cE_y/B$ as if the substrate potenial $V(x,y)$ was absent [17,18]. As a consequence, *a state with a Hall velocity greater than $cE_y/B$* cannot be in the sudden approximation, i. e., it *must partially follow an adiabatic time evolution*. This means that in the quantum Hall regime the velocities of the conducting states contain non-linear components. This conclusion is important for the theory of the quantum Hall effect.

Let us note here again that in the absence of any substrate potential all Hall velocities are equal to the classical value $cE_y/B$. It needs the presence of a substrate potential to generate lower or higher values of the Schrödinger Hall velocities. Further, the Hall velocity of any linear combination of unperturbed Landau functions *of a single band* is still equal to $cE_y/B$ whenever the coefficients are time-independent. Therefore it can be different from this value only with time dependent coefficients. With our boundary conditions this leads to terms arising from particle densitiy redistributions among Landau functions located at different positions in x-direction. (In the model calculations [15,5,16] these density redistributions are generated by scattering (exchange) between functions localized on adjacent opposite slopes of the smooth part of the disorder potential.)

We would like to finish this section by emphasizing that for the explanation of quantized Hall *plateaus* it is not sufficient to understand the existence of localized (i.e., insulating) states in the tails of the broadened Landau bands. (Actually, on a pure theoretical ground such states are not indispensable for the quantization of the *Hall* conductivity since in very special situations the Hall conductivity is quantized in the absence of insulating states and the entire compensating current is generated by electron-phonon interaction [19]). Rather one has to understand in what situations the sum of all the Hall velocities leads to $\sigma_{xy} = (e^2/h) \times integer$, that is, one has to understand the microscopic mechanism of the compensating current. This includes in particular the understanding of the mechanism which leads to Hall velocities greater than $cE_y/B$ for certain filling factors. This mechanism and that of the resulting compensating current has been understood in detail for model systems of the integer quantum Hall effect [15,5,16,20].

## 5. TWO KINDS OF NON-LINEARITIES

We have seen that the total current is the sum of velocities which in general are composed of one or several *adiabatic* parts times the corresponding probabilities for adiabatic evolution and of one or several *non-adiabatic* velocity parts, which include the probabilities for the corresponding non-adiabatic transitions. In general these probabilites depend themselves on $E_y$ in a non-linear way. The non-adiabatic transition probabilities (within the interval $\tau/2$) tend to zero for sufficiently low $E_y$ and saturate to a constant value for sufficiently high $E_y$. This means that the corresponding transition rates (averaged over this time interval) may not always be quadratic in $E_y$, for instance close to a mobility edge, i. e., for states which are near the adiabatic limit for the given value of $E_y$ or for states close to the sudden approximation. In addition, adiabatic velocities are intrinsically non-linear due to their periodicity in $E_y$ with period $|h/(eL_y t)|$ for a given t, as we have seen. This means that, on a microscopic level, the current can be non-linear as a result of *these two types of non-linearities.*

Futher, since the time averaged transition probabilities are in general non-linear in $E_y$, even the *macroscopic*, i. e. time averaged current can be non-linear in certain ranges of $E_y$. This occurs in particular if the Fermi level is near a mobility edge. In this case, by changing the strength of the electric field $E_y$, the mobility edge can be shifted across the Fermi level and the *parallel conductivity* of the system changes from zero to non-zero and vice versa. Such a situation occurs in a quantum Hall system, where increasing the macroscopic electric field shifts the mobility edges towards the edges of the broadened Landau bands, by this causing a shrinking of the widths of the quantized plateaus (see [18]). This macroscopic non-linearity is thus a direct consequence of the microscopic non-linearities. For sufficiently high electric fields no adiabatic states remain in the bands which means that the plateaus of zero parallel conductivity have shrunk to zero (dissipative breakdown of the quantum Hall effect).

The macroscopic *Hall current* is the sum over the (time averaged) Hall velocities of all occupied states. When the Fermi energy remains in a range of adiabatic states (and in addition, when an intrinsic condition for the sudden approximation states in the band center is fulfilled, see [21], sec. 5, conditon 2) this sum has the value $I_x = E_y(e^2/h) \times integer$, i. e., the Hall conductivity is quantized and the *macroscopic* Hall current is *linear* in $E_y$ in spite of *non-linear microscopic* contributions.

## 6. SUMMARY

We have investigated the Schrödinger time evolution of charged particles in an asymmetric substrate potential (disorder) and a homogeneous macroscopic electric field and shown that it is in general non-linear with respect to the field. In particular, the microscopic velocities contain non-linearities of two kinds. The first kind arises from the highly non-linear adiabatic evolution and (or) from an admixture of parts of it in so-called intermediate states, and the second kind from the generally non-linear transition probabilities between adiabatic states.

The resulting macroscopic dc-current may or may not be linear in the field. Three cases can be distinguished. In the first case the microscopic non-linearities are not relevant for the macroscopic current. This is the scenario of linear response theory (Kubo formalism), which is based on linearising the Schrödinger time evolution with respect to the macroscopic electric field, i. e., which neglects the microscopic non-linearities. We have given arguments which make it plausible, that in many cases such a theory is justified for the current parallel to the electric field.

In the second case the microscopic non-linearitites also lead to macroscopic non-linearities. As an example we mentioned the onset of dissipation by increasng the electric field in breakdown phenomena of the quantum Hall effect.

In the third case the macroscopic current is linear although the microscopic non-linearities cannot be neglected and constitute an essential part of it. An example is the Hall current in the quantum Hall regime. This shows that macroscopic linearity does not necessarily imply microscopic linearity.

Further, we discussed some other questions raised by the Kubo theory. We elucidated the fact that this formalism leads to zero parallel conductivity if applied to a system with discrete levels. This is shown to be an unphysical restriction which arises from the neglect of the full (non-linear) time evolution of the adiabatic basis states. If this evolution is taken into account, one obtains in general a non-zero value also in the case of discrete levels. Incidentally, if then the limit towards a continuum is taken, we recover, but only under certain conditions, the linear expression (based on Fermi's golden rule) for the parallel conductivity.

The existence of a parallel conductivity implies dissipation. But in linear response theory a mechanism of dissipation (interaction with the phonons of a heat bath) and the calculation of the correponding steady state are absent. Instead, the unperturbed Fermi distribution appears in the final expressions, i. e., the distribution before the electric field has been applied. This is a consequence of the linearization and of the assumption that the deviation of the steady state from the original distribution is sufficiently small. But the existence of a steady state must then be *postulated*. In this sense linear response theory is phenomenological [14]).

## APPENDIX A

We will show that the transition rates $dp_{nm}/dt$ derived from expression (8) formally lead to Fermi's golden rule under the following conditions : $\alpha_{mn} \approx$ const. , $(\varepsilon_m - \varepsilon_n)/\hbar = \omega_{mn} \approx$ const.. With the last two conditions it is shown in Ref. [9], page 753, that the probability to go from $u_n(0)$ to $u_m(t)$ can be written in the form

$$p_{nm} \approx |\alpha_{mn}/\omega_{mn}|^2 \, 2[1 - \cos(\omega_{mn}t)] \tag{A1}$$

Further, from ref. [9], page 761, exercise 6, one gets

$$\alpha_{mn} = -<m|dH/dt|n>/[\hbar \omega_{mn}(t)] \tag{A2}$$

With $dH/dt = eE_y v_y^{op}$ and since $\alpha_{mn}$ and $\omega_{mn}$ are constant by hypothesis, we can write $\alpha_{mn} = eE_y <m|v_y^{op}(t=0)|n>/[\hbar \omega_{mn}]$. Further, expressing in (A1) the factor $2[1 - \cos(\omega_{mn}t)]/(\omega_{mn})^2$ in the usual way as $2\pi t \delta(\omega_{mn})]$, see e. g. [9] (XVII.43), which is valid for sufficiently large t, one obtains for the transition rate $dp_{nm}/dt \approx |eE_y<m|v_y^{op}(t=0)|n>|^2 2\pi\delta(\omega_{mn})/[\hbar^2(\omega_{mn})^2]$, which is Fermi's golden rule.

## APPENDIX B

We will give a proof of the non-linearity of the adiabatic evolution making use of general properties of the momentary eigenfunctions $u_n(\mathbf{r};\phi)$ of the Hamiltonian (1), $H(\phi)$, were $\phi = -cE_y L_y t$ is considered as a parameter. Hamiltonians of this form have been studied in a general way in [8] for systems with discrete, non-intersecting energy levels $\varepsilon_n(\phi)$. We have already mentioned that the levels $\varepsilon_n(\phi)$ are periodic in $\phi$ with period $|hc/e|$, which corresponds to a periodicity in t with period $\tau = h/|eL_y E_y|$ and to a periodicity in $E_y$ with period $h/|eL_yt|$. As a consequence of (4), this implies the periodicity of the *adiabatic velocities parallel* to $\mathbf{E} = (0,E_y,0)$ and hence their non-linearity with respect to t and $E_y$, provided the levels are not flat.

Concerning the *Hall velocities* $v_x$, we make use of the result [8] that also the modulus $R_n(\mathbf{r};\phi)$ of an eigen function $u_n(\mathbf{r};\phi) = R_n(\mathbf{r};\phi)\exp[i\theta_n(\mathbf{r};\phi)]$ is periodic in $\phi$ with period $hc/e$. This leads to an oscillation of the particle density $[R_n(\mathbf{r};\phi)]^2$ as a function of $\phi$, *provided* $R_n(\mathbf{r};\phi)$ does *not remain constant* when $\phi$ changes in the interval $hc/e$.

That $R_n(\mathbf{r};\phi)$ does not remain constant, can be seen as follows (we drop the index n). Consider the phase winding numbers

$$W_P = 1/(2\pi) \int_P \mathrm{grad}\theta \, d\mathbf{r} \qquad (B1)$$

where P is any fixed path which crosses the systems domain in y direction from (x;0,z) to $(x,L_y,z)$. The numbers $W_P$ change with $\phi$ according to the general law [8]

$$W_P(\phi + hc/e) = W_P(\phi) + 1. \qquad (B2)$$

Relation (B2) implies that, during the $\phi$–interval $hc/e$, one or several phase vortex lines, which are the center of a multi-valued phase function, cross the path P and thus traverse the physical domain. Since these lines are nodal lines (where $u(\mathbf{r};\phi)$ vanishes) this implies a genuine $\phi$-dependence of $R(\mathbf{r};\phi)$.

In general this leads to an oscillation of the center of gravity of the particle density, hence to a non-vanishing, oscillating contribution $\partial_t \int [R(\mathbf{r};\phi)]^2 \mathbf{r} dV$ to the particle velocity, which in general has a non-vanishing component in x-direction, which is periodic in $\phi = -cE_yL_yt$ and hence non-linear in time and in $E_y$. It is straight forward to extend this result to adiabatic many particle states since the winding number law (B2) also holds for the many-particle case [8]. An explicit example for the law of winding number change (B2) in the presence of a magnetic field and the associated adiabatic particle density and Hall velocity oscillation is given in [22].

24**FIGURE CAPTIONS**

Fig. 1  Schematic representation of adiabatic energy levels as a function of time. Dashed lines with full arrow heads indicate possible non-adiabatic transitions from one level to other levels, the open arrow head symbolizes the adiabatic evolution of this level. t is the time in units of $\tau$ (see text).

Fig. 2  Energy levels of a one dimensional system of length L (a) and of length 2L (b), see text. Dashed lines show the levels $E_k$, $E_q$, ... in the presence of a parallel electric field F. Full lines show the adiabatic levels when in addition a small disorder substrate potential is present (weak disorder case). The $\Delta$'s indicate the length of the anti-crossing intervals, e. g., $\Delta_{kk'}$ indicates the transition time during which the adiabatic function $u_n(t)$ of Fig a) or $u'_m(t)$ of Fig. b) change from the plane wave $\psi_k$ to the plane wave $\psi_{k'}$. t is the time in units of $\tau = h/|(eLF)|$.

Fig. 3a and b qualitatively show the absolute values of the expressions $\alpha_{n+1n}$ and $\alpha_{m+1m}$ associated with the non-adiabatic transitions from $\varepsilon_n$ to $\varepsilon_{n+1}$ in Fig. 2a and from $\varepsilon'_m$ to $\varepsilon'_{m+1}$ in Fig. 2b.

Fig. 4  Time evolution in a weak disorder scenario. The numbers indicate the evolution of the initial function $u_i(t=0)$, starting at point 1, developing into a linear combination of $u_n(t)$, see text. t is the time in units of $\tau$.

Fig. 5  Schematic illustration of the time evolution of the conducting states in a broadened Landau band. These states undergo non-adiabatic transitions (schematically indicated by dashed lines) with probabilities which decrease from the band center towards the two mobility edges, see text. Shown are the first few conducting states close to the upper ($\varepsilon^+$) and lower ($\varepsilon^-$) mobility edge. Beyond the mobility edges the states follow the adiabatic evolution and do not undergo non-adiabatic transitions.

Fig. 1

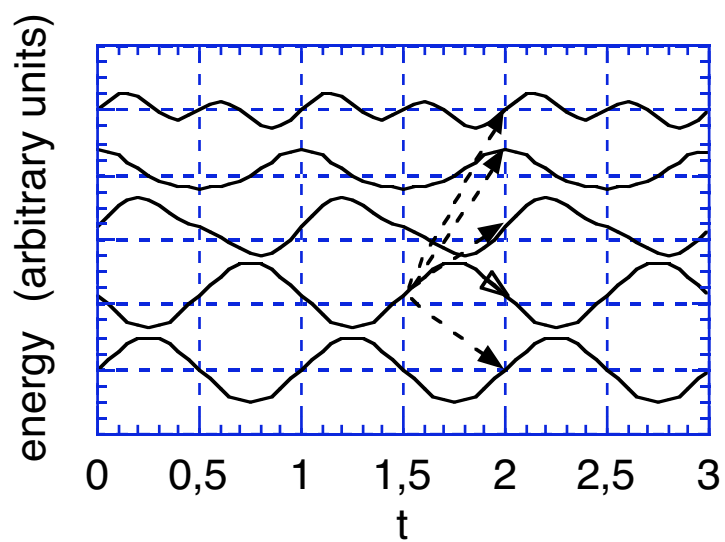

Fig. 2a

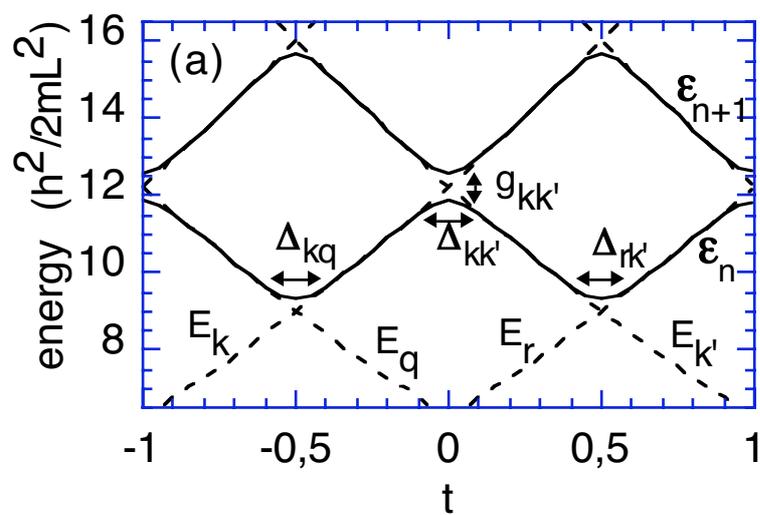

Fig. 2b

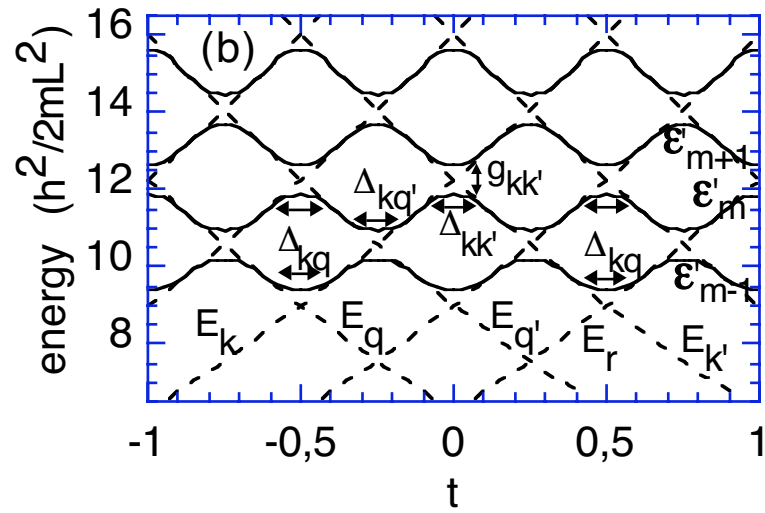

Fig. 3ab

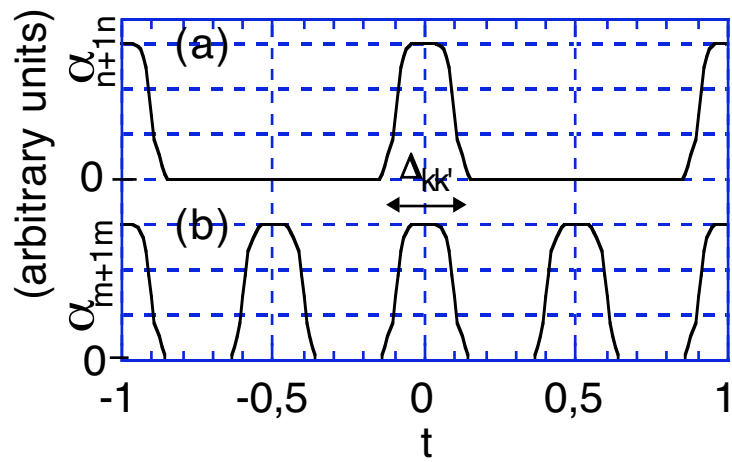

Fig. 4

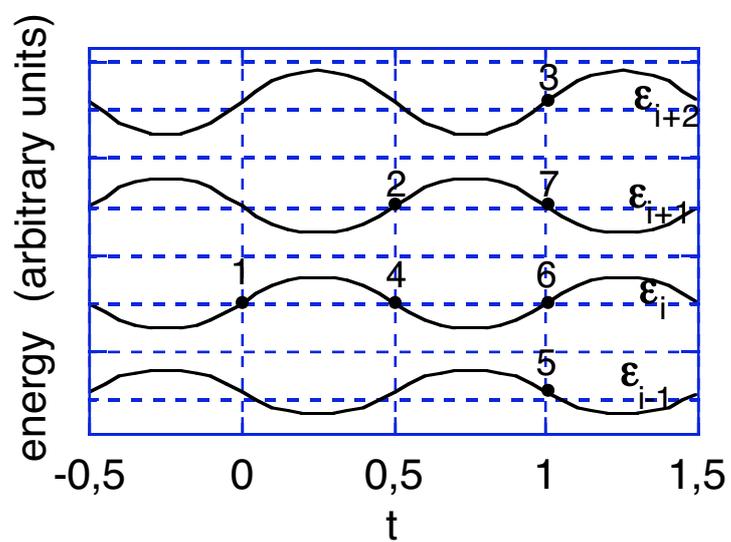

Fig. 5

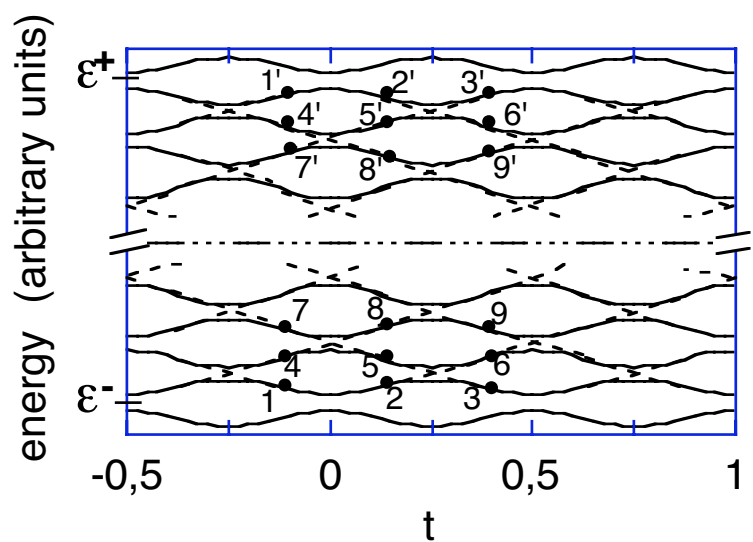